\documentclass{aa}
\usepackage{graphics,times}
\begin{document}

   \thesaurus{10.
              (02.18.5;
               10.03.1;  
               13.18.2;
               13.25.2)} 
   \title{Synchrotron radiation from quasi-monoenergetic electrons}

   \subtitle{Modelling the spectrum of Sgr A$^*$}

   \author{T. Beckert  \inst{1}
           \and
	   W.J. Duschl  \inst{1,2}
          }

   \offprints{T. Beckert 
              (tbec@ita.uni-heidelberg.de)
              }

   \institute{Institut f\"ur Theoretische
              Astrophysik der Universit\"at Heidelberg, Tiergartenstr. 15,
              D-69121 Heidelberg, Germany
	      \and
	      Max-Planck-Institut f\"ur Radioastronomie, Auf dem H\"ugel 69,
	      D-53121 Bonn, Germany
             }

   \date{Received ...; accepted ...}

   \maketitle

\begin{abstract}
We investigate the spectrum of a quasi-mono\-en\-er\-ge\-tic ensemble of
relativistic electrons, especially for the mildly relativistic case, and
discuss the effect of inclination of the magnetic field on the emissivity.
We apply the exact
theoretical description to the spectrum of the radio source Sgr A* which is 
located at or very close to the dynamical center of our Galaxy. We find that 
the radio-MIR spectrum can be reproduced well, but that the resulting
self-comptonized X-ray flux is much smaller than the observed one. 
\end{abstract}
      
\section{Introduction}
Observations of the galactic center radio source Sgr A$^*$ in the last 
few years (e.g. Zylka et al.\ 1995) have raised the interest in 
detailed investigations of synchrotron spectra produced by 
mildly relativistic electrons (e.g. Melia 1994, Mahadevan et al. 1996). 
 \cite{Mel94} proposed synchrotron radiation from thermal electrons 
in a radial inflow but failed to explain the latest sub-mm data. 
Narayan et al. (1995) presented a model for the entire
spectral range from radio to X-ray frequencies based on 
advection dominated accretion. 

In our view, only the radio observations can be taken as firm 
detections and the best fit from accretion models with LTE 
is based on a isothermal sphere emitting 
optically thin synchrotron radiation. In a previous  paper (Beckert et al.\ 
1996, hereafter referred to as  B96)  we showed that a 
homogeneous blob of relativistic electrons penetrated by a magnetic field is
sufficient to explain the observed spectrum. 

In the present paper we describe the calculation of synchrotron spectra of 
electrons with various distributions in momentum space as seen by an arbitrary 
observer. The basic equations are presented in Sect.\ \ref{theo} and  
Appendix A. We apply the general formulae (e.g. \cite{Bek66}) without
approximations to a source of mildly to highly relativistic electrons. 
In Sect.\ \ref{mol} we discuss in detail the synchrotron spectrum of 
monoenergetic electrons and compare the results with the standard treatment for
highly relativistic electrons. 
In Sect.\ \ref{quas} we provide a criterion for quasi-monoenergetic 
distributions and show the deviations from a simple approximation to
the synchrotron spectra which is used to identify optically thin monoenergetic
synchrotron sources (Duschl \& Lesch 1994, Reuter \& Lesch 1996). In the 
subsequent Sect.\ \ref{flu} we derive the synchrotron flux of 
homogeneous spherical sources with optical thick-thin transition. 
Sect.\ \ref{SGR} deals with the application of
quasi-monoenergetic electron distributions to the observations of Sgr A*. 
We interpret the low-frequency turn-over (B96) in terms of synchrotron 
self-absorption
(SSA) and determine the physical source parameters under the assumption
of energy equipartition between electrons and magnetic field. An 
unavoidable consequence of the synchrotron mo\-del is the self-comptonization of 
the emitted radiation (SSC) which emerges as X-rays, 
briefly discussed in Sect.\ \ref{ssc}. 
We summarize our results in Sect.\ \ref{conc}. 
\section{Synchrotron Theory \label{theo}} 
The theory of synchrotron emission has been reviewed
several times (\cite{Bek66}, Ginzburg \& Syrovatsky 1969).
Here we briefly discuss the basic formulae for the coefficient of spontaneous
emission and for the observed power of synchrotron radiation of single
electrons. The fundamental work was
done by Schwinger (1949) and Westfold (1959), but later on the difference
in observed and emitted power of single electrons was recognized by 
Epstein \& Feldman (1967) and Scheuer (1968).   
We use a formulation of synchrotron emision based on the retarded potentials
to calculate the recieved power seen by a distant observer. Radiation losses 
are neglected and we assume that binary encounters do not take place.
The relativistic electrons are characterised by their 
energy $E = \gamma m_e c^2$, the orbital frequency
   \begin{equation}
   \omega_s = \frac{qB}{\gamma m_ec} 
   \qquad
   \gamma = \frac{1}{\sqrt{1-\left|\vec{\beta}\right|^2}}
   \end{equation}
in a locally homogeneous magnetic field $\vec{B}$ and a pitch angle $\alpha$
between the magnetic field and the momentum of the spiraling electron. 
 The electric and magnetic radiation field 
which determine the flux density are peroidic in the retarded time of emission $\tau$
   \begin{equation}
   t = \tau + \frac{R(\tau)}{c} \qquad .
   \end{equation} 
Rather than performing a Fourier transformation to get the spectral
emissivity of a single particle in a homogeneous magnetic field we
expand the radiation field in a Fourier series. The recieved power 
per steradian in a direction $\theta$ to the magnetic field is derived in
Appendix A and reads  
   \begin{equation} \label{grom}
   P = \frac{q^2}{2\pi c} 
   \sum_{m=1}^\infty \left|\vec{g}_m\right|^2
   \end{equation}
with a complex field vector
   \begin{equation}\label{sqr}
   \vec{g}_m  =  \frac{m\omega_s}{ \xi^2}
   \left(\begin{array}{c}
   \beta\sin\alpha\,J'_m(m\psi)\\
   i(\beta\cos\alpha - \cos\theta)\sin\theta^{-1}\,J_m(m\psi)  \\ 0 
   \end{array} \right)
   \end{equation}
depending on the geometric quantities 
   \begin{equation} \label{freq}
   \psi = \frac{\beta\sin\alpha\sin\theta}{\xi} 
   \qquad \xi = 1 - \beta\cos\alpha\cos\theta \qquad .
   \end{equation}
The coefficient of spontaneous emission $\eta_\nu$
which is related to the emissivity by 
$j_\nu = \int {\rm d}^3p\;\xi\,f(p)\,\eta_\nu(p)$
is obtained with a  
resonance condition $ \delta\left(\nu - m\nu_s \xi^{-1}\right)$, since
we are dealing with emission at discrete frequencies $\nu = m\nu_s \xi^{-1}$.
\begin{eqnarray} \label{emis}
\eta_\nu & = & \frac{q^2 (2\pi \nu)^2 \beta^2}{2 \pi c \xi^2} \sum_{m>1}
\delta\left(\nu - m\nu_s \xi^{-1}\right) \times \nonumber \\ 
& & \left[ \left(\sin\alpha J'_m(m\psi)\right)^2 + 
\left(g_1 J_m(m\psi)\right)^2 \right] \\
   g_1 & = & \left(\frac{\beta\cos\alpha_m 
    -\cos\theta}{\beta\sin\theta}\right) \label{ge1}  \qquad .
\end{eqnarray}
This form of the spontaneous emission differs from the standard form
(e.g. \cite{Bek66})
by an additional factor $\xi^{-1}$. In the standard calculation a square of 
$\delta$-functions appears. From our point of view, one of these 
is replaced by $\omega_s^{-1}$ and divided by $T = 2\pi\omega_s^{-1}$
to get the mean power from the total energy. But the frequncy $\nu_s = T^{-1}$ 
is not the observed frequency $\nu = \xi \nu_s$ of the radiation. 
This additional $\xi$ is
contained in our Eq. (\ref{emis}). We will in see Sec. \ref{mol} that
this difference is offset by the
normalization of the distribution function in calculating the spectral 
emissivity.  
\section{Synchrotron spectra of monoenergetic electrons \label{mol}}
   \begin{figure}[htbp]
      \resizebox{\hsize}{!}{\includegraphics{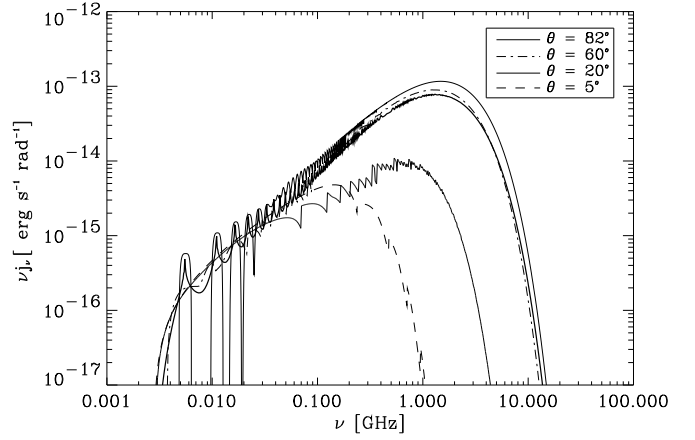}}
      \caption{The spectrum of mildly relativistic electrons 
               with $\gamma = 5$ seen by observers 
               at $\theta = 5\degr, 20\degr, 60\degr, 
               82\degr$. We plot the received power $\nu j_\nu$ for a
               normalized distribution.  The
               spectra obtain their maxima at higher frequencies as 
               observers approach $\theta = 90\degr$.
               The thickest solid line is the average of all observers 
               corresponding to an isotropic
               distribution of magnetic field directions.}
         \label{mfs}
   \end{figure}

   \begin{figure}[htbp]
      \resizebox{\hsize}{!}{\includegraphics{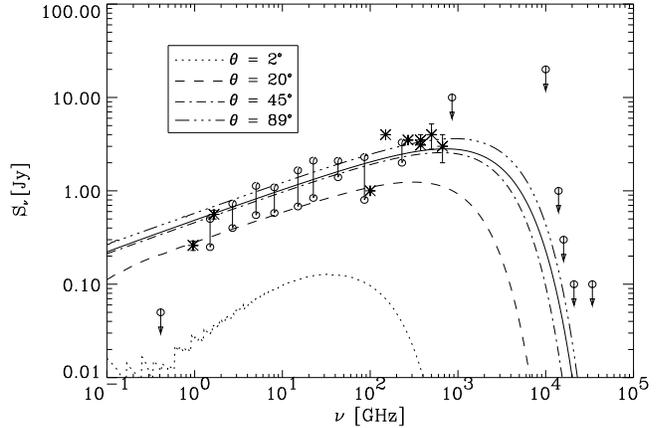}}
      \caption{The flux density distribution of Sgr A$^*$ 
               fitted with monoenergetic electrons of $140$ MeV 
               corresponding to $\gamma = 275 $. The
               observed spectrum is taken from B96 where symbols
               with circles at both ends indicate variability.
               The solid curve corresponds to the mean observer position and
               broken lines refer to $\theta = 89\degr, 45\degr,
               20\degr, 2\degr $ from
               high to low flux densities at $10$ GHz. For physical 
               parameters, see model {\sc i} in Tab.\protect\ref{smy}.}
         \label{mono}
   \end{figure}
We describe the radiating electrons by a stationary distribution function 
$f_e(\alpha,p)$ in momentum space in a homogeneous magnetic field. 
The distribution in phase 
space is assumed to be separable and the complete distribution function 
looks like $n(\vec{r}) f_e(\alpha,p)$ with the volume density $n(\vec{r})$.
In order to recover the standard description of emissivity we must consider
the distribution $f_e$ in the observers frame.  We take $f_e$
to be normalized so that the
density of electrons in the source is contained in $n(\vec{r})$. 
In principle, the distribution
is a density and can be time dependent. While the distribution describes the 
electrons in the source at time $\tau$, we have changed the time variable to the
time of observation $t$. This transformation is reflected in the differential
d$t = \xi$ d$\tau$ and the distribution seen by the observer is therefore $\xi 
f_e(\alpha,p)$. A more extended discussion of this argument is given in 
Appendix B. 
The integrated spectral emissivity in the observers frame
follows from the power of a single electron stated above
   \begin{equation}
   j = \frac{q^2}{c}\int_0^\infty {\rm d}p
   \;p^2 \int_0^\pi{\rm d}\alpha \,\sin\alpha\,\xi\,f_e(\alpha,p)
   \sum_{m=1}^\infty \left|\vec{g}_m\right|^2 \qquad .
   \end{equation}
The spectral emissivity is 
obtained from the spontaneous emission given in Eq. (\ref{emis})
by evaluating the $\alpha$-integration. The $\delta$-function
fixes $\alpha$ for each combination of $[\nu,\theta \neq \frac\pi 2,m ]$ 
unambiguously by
   \begin{equation} \label{alp}
   \cos\alpha_m = \frac{1-\frac{m\nu_s}{\nu}}{\beta\cos\theta} \qquad .
   \end{equation}  	
Since $\alpha_m \in [0,\pi]$ has to be a real number, we get a condition
for the index $m$ 
   \begin{equation}
   \frac{\nu}{\nu_s}(1-\beta|\cos\theta|) \leq m \leq 
   \frac{\nu}{\nu_s}(1+\beta|\cos\theta|) \qquad .
   \end{equation}
It turns out that the problem of deducing the spectrum of a distribution
of electrons in momentum space is reduced to one integral and a limited sum 
of Fourier coefficients :
   \begin{eqnarray}
   \label{pf}
   j_\nu(\Omega) & = & n(r)
   \frac{4\pi^2 q^2\nu}{c}\int_0^\infty  {\rm d}p
   \;p^2 \sum_m \frac{\beta\,f_e(\alpha_m,p)}{|\cos\theta|} \times
   \nonumber   \\ & &
   \left[ \left(\sin\alpha_m\,J'_m(m\psi)\right)^2 + 
   \left(g_1\,J_m(m\psi)\right)^2\right] \quad .
   \end{eqnarray}
The abbreviation $g_1$ appeared previously in Eq. (\ref{ge1}).
This form is an exact result with $\alpha_m$ determined by Eq. (\ref{alp}).
The sum can be evaluated directly for
small frequencies $\nu \sim \nu_s$ and small momenta where only a few
cyclotron lines emerge in the spectrum. For frequencies where the range of 
indicies $m$ for the harmonics contributing to the sum is wide, 
it is reasonable to transform the sum to an integral and evaluate 
it numerically.

For a monoenergetic and isotropic distribution, the distribution
function is $f_e(\alpha_m,p) = (4\pi p_0^2)^{-1}\;\delta(p-p_0)$ and 
the emissivity in Eq. (\ref{pf}) consists of contributions from electrons
described by $(\alpha_m, p_0)$ whose $m$-th harmonic is shifted to 
the frequency $\nu$ in the direction of the observer.
The resulting spectrum is shown for four different observers in Fig.\ref{mfs},
for electrons with a common Lorentz factor of $\gamma = 5$ 
corresponding to $2.56$ MeV. We see that the average of all observers is 
similiar to the emission in the direction $\theta = 60\degr$ shown by the 
dashed line in the Figure. 
The received flux changes drastically when the direction to the observer becomes
parallel to the magnetic field lines. The emission for $\theta = 5\degr$
consists of overlaping low $m$-harmonics of electrons with small pitch angels. 
In addition, it demonstrates the transition from overlaping cyclotron
lines to a continuous synchrotron spectrum. 
The flux measured by the observer is increased relative to
the emitted power of the spiraling electron which mainly radiates in the
direction parallel to its momentum. Ginzburg et al. (1968) 
recognized that there also is a change 
in the total energy of the radiation field inside a fixed sphere 
in the observers frame containing the electron and
the observer at its surface.  

The synchrotron spectra of monoenergetic electrons provide a reasonable fit
to the Sgr A$^*$ spectrum except for the turn-over below 1 Ghz. 
We have taken the average of all observer positions
and a magnetic field strength of $10$ G, comparable to the first model 
using monoenergetic electrons by Duschl \& Lesch (1994).
The $\theta$-dependency is increased for high
energies as seen in Fig.\ref{mono}.  The parameters are combined
in Tab.\ref{smy} as model {\sc i} and discussed in greater detail in Sect.\ 
\ref{SGR}. 

\begin{figure}[htbp]
      \resizebox{\hsize}{!}{\includegraphics{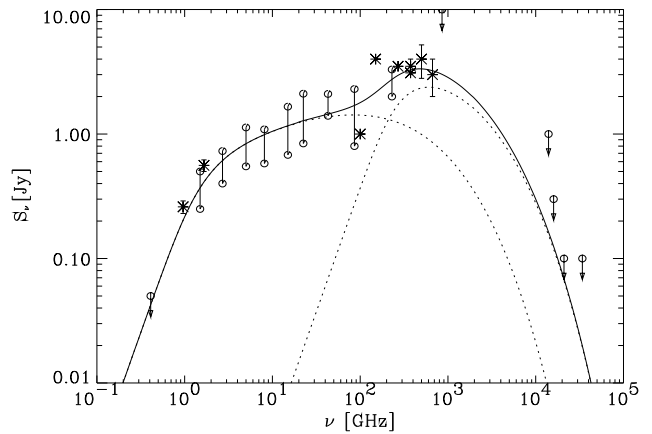}}
      \caption{The spectrum of a core-shell model of Sgr A$^*$ consisting
               of two thermal emitting regions which are homogeneous in itself.
               The model parameters are given in Tab.\protect\ref{com} for
               the extended shell as part {\sc i} and the self-absorbed 
               compact core {\sc ii}. Both components are drawn separately
               as dotted lines.
              }
         \label{two}
   \end{figure}

\begin{figure}[htbp]
      \resizebox{\hsize}{!}{\includegraphics{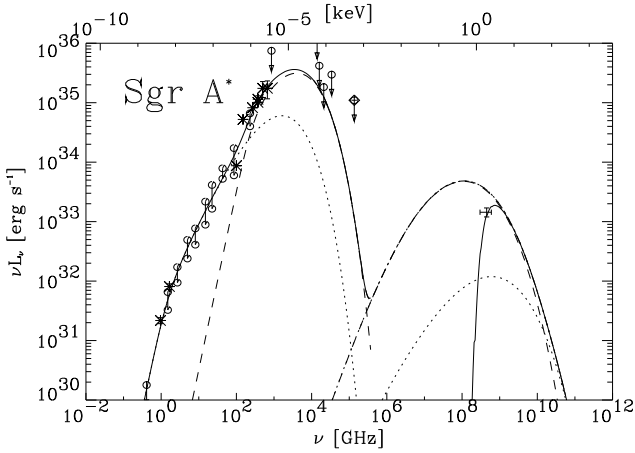}}
      \caption{The luminosity of Sgr A$^*$ from the radio to the soft X-ray 
               regime for an assumed distance of 8.5 kpc.
               The ROSAT flux at 1.85 keV is included.
               The two-component synchrotron spectrum 
               from a hot ($T = 6.0\ 10^{11}$ K) extended envelope {\sc i}
               and a cooler ($T= 1.6\ 10^{11}$ K) compact core {\sc ii} 
               (the same as in Fig.\ \protect\ref{two}) is
               fitted as a solid line. It includes SSC and
               absorption as discussed in Sec. \protect\ref{ssc}. The 
               intrinsic spectrum of the core (envelope) 
               corresponds to the dashed (dotted) line. The physical parameters
               are given in Tab.\protect\ref{com}.              }
         \label{poe2}
   \end{figure}

So far we have a description for the spontaneous emission. To get a complete
spectrum, we must include induced emission and absorption inside the source.
The cross section for synchrotron self-absorption (SSA) is basically given by 
   \begin{equation} \label{cross}
   \sigma_\nu  =  \frac{c^2}{8\pi h \nu^3}\int_0^\infty {\rm d}p p^2 
   (f(p_1)- f(p))\;\epsilon_\nu(p)
   \end{equation}
where $p_1$ is the momentum of the electron before absorption. 
The relativistic limit - which we assume for SSA - is simply $p_1c = pc - h\nu$.
The emissivity $\epsilon_\nu(p)$ is the $\theta$-average of $j_\nu$
per electron.
In principle the emissivity has to be split up into the two directions of
polarisation present in Eq. (\ref{sqr}) and the cross-section must be 
calculated 
separately for both directions. Since we assume that the distribution of
magnetic field directions is isotropic on a length scale much shorter than
the source size, this distinction can be omitted without indroducing relevant 
errors.
Using a Taylor-expansion for $ f(p_1)- f(p)$ we get an 
expression which closely resembles the standard formula and can easily 
be evaluated numerically.
   \begin{equation} 
   \sigma_\nu = \frac{c^2}{8\pi h \nu^3}\int_0^\infty 
   {\rm d}p (p_1-p)\frac{\partial}{\partial p} f \sum_m
   \frac{\beta \nu }{m\nu_s\left|\cos\theta\right|}\left|\vec{g}_m
   \right|^2 \; .
   \end{equation}
All low frequency turn-overs seen in the spectra of 
Fig. \ref{therm} - \ref{comp} are due to sychrotron self-absorption.
\section{A criterion for quasi-monoenergetic distributions \label{quas}}
\begin{figure}[htbp]
      \resizebox{\hsize}{!}{\includegraphics{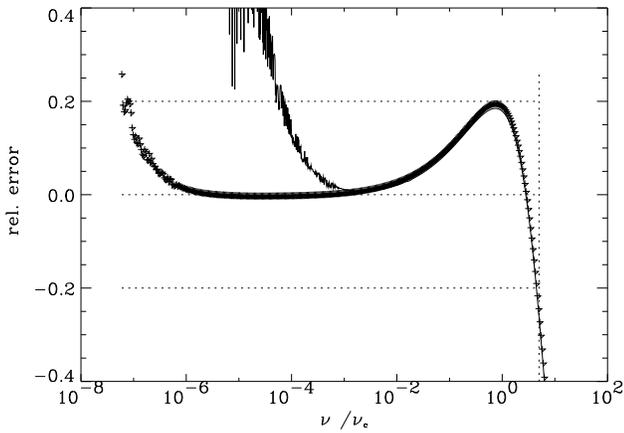}}
      \caption{Relative error of the simple approximation 
        Eq.(\protect\ref{smpl})
        compared with the monoenergetic case averaged over the observer
        positions $\theta$. The crosses indicate the error for 
        $\gamma = 1000$ and the solid line for $\gamma = 100$. The deviations
        at small frequencies are dominated by the discrete cyclotron lines
        present in the strict monoenergetic case. The magnetic field is taken as
         $B = 10$ G. }
         \label{appr}
   \end{figure}

Untill now we have not defined what we call a quasi-mo\-no\-en\-er\-ge\-tic 
electron distribution. To do this we compare the exact formulae given in 
Sec. \ref{mol} with a simple approximation for the emissivity of 
monoenergetic electrons in the energy range $\gamma = 100 \ldots 1000$. 
This will allow us to derive analytic expressions for the
emissivity of truncated power-law distributions. From these we get a criterion
what a quasi-monoenergetic distribution will be when synchrotron emission 
is considered.   
We approximate the synchrotron spectrum of a single electron by 
  \begin{equation} \label{smpl}
  P(\nu,\gamma) = P_0 \nu^{1/3} \exp[-\nu/\nu_c] 
  \end{equation}
which is used in the interpretation of the radio spectra of Sgr A$^*$ (Duschl
\& Lesch 1994) and the core of M81 (NGC3031) by Reuter \& Lesch (1996).
The total luminosity of a single electron is
\[ L = L_0 B^2 (\gamma\beta\,\sin\alpha)^2 \]
with pitch angle $\alpha$ in the frame of the source and
\[ L_0 = 1.5870\,10^{-15} {\rm erg}\;{\rm s}^{-1}\]
if the magnetic field is measured in Gauss. In the observers frame the shift 
of time intervals d$t = \xi$d$\tau$ introduces a factor $\xi^{-1}$ as 
discussed in Sec. \ref{mol} and in Appendix B. 
For sufficently high electron energies and
frequencies much larger than the orbital frequency, the radiation is beamed
in the direction $\alpha \approx \theta$ and we approximate $\xi \approx
\sin^2\alpha$. The observed luminosity becomes independent of the pitch angle :
\[ L = L_0 B^2 \gamma^2 \]
The scaling of the spectral power $P(\nu,\gamma)$ is obtained 
from the luminosity and gives
\begin{equation}\label{cri} 
  \nu_c = \nu_0 \gamma^2 = \frac{\kappa B q \gamma^2}{2\pi m c} 
\end{equation}
\[ P_0 = \frac{3}{\Gamma(1/3)} L_0 B^{2/3} \gamma^{-2/3} 
\left(\frac{ 2\pi m c}{\kappa q} \right)^{4/3} \qquad. \]
As a factor in the critical frequency $\nu_c$ we introduced a coefficient $\kappa$
which is usually taken as $\kappa = 1.5$. We would like to compare the
approximation with the exactly evaluated spectra and minimize the relative
error. This suggests a smaller value for $\kappa \approx 1.2$ so that the
error is less than 20\% in the range $100 < \gamma < 1000 $ as shown in Fig.
\ref{appr}. We restrict the error box in Fig. \ref{appr} to frequencies around 
the maximum of the flux. At high frequencies $\nu > 5 \nu_c$ the spectrum drops
nearly exponentially and the flux has decreased by two orders of magnitude.
At low frequencies the spectrum is changed by the presence of cyclotron lines
which provide a significantly smaller flux. But the absolute flux is 
small at low frequencies due to the approximated $P_\nu \sim \nu^{1/3}$ 
behavior and is often suppressed by self-absorption.  
In order to get analytic expressions which determine what we want to call a
quasi-monoenergetic electron distribution, we investigate
restricted power-laws  $ f_0 \gamma^{-\alpha}$
for $\gamma_1 < \gamma < \gamma_2$ and set the distribution to zero 
for other energies. The width $\chi$ of the distribution is defined as 
$\chi = \gamma_2/\gamma_1$ and is the main parameter in our investigation. 
The mean luminosity is then
\[ \left< L \right> = n(r)L_0 B^2 \left< \gamma\,\right>^2\,\frac{Q^{2}(2)}{Q(1)\,Q(3)}\] 
\[ 
  Q(x) = \left(\frac{x-\alpha}{\chi^{x-\alpha}-1}\right)
\] 
We normalize the distribution function so that
$n(r)$ gives the spatial density of the relativistic electrons.
\[ n(r,\gamma) = n(r) f_0 \gamma^{-\alpha}
 \]
\[ f_0 = \gamma_1^{\alpha -1}\,Q(1)\qquad\left<\gamma\,\right> = \gamma_1 \frac{Q(1)}{Q(2)} \]
We assume equipartition between magnetic and kinetic energy density. 
Despite the presence of a thermal plasma, we only consider the 
relativistic electrons in the kinetic energy. The spatial density 
\[  n(r) = \frac{B^2}{8\pi m c^2}
\left<\gamma\right>^{-1} \]
is thus unambiguously determined by the magnetic field and the
mean electron energy.
In the discussion of quasi-mo\-no\-en\-er\-ge\-tic distributions, we want to redefine the
critical frequency given in Eq.(\ref{cri}) by replacing the square of the
relativistic $\gamma$-factor by the square of the mean $\left<\gamma\,\right>^2$
\[ \nu_c = \frac{\kappa B q \left<\gamma\,\right>^2}{2\pi m c} \qquad .\]
The mean emissivity is 
\begin{equation}\label{mem}
 j_\nu = j_0 B^{8/3} \nu^{1/3}\left<\gamma\right>^{-\frac{5}{3}}\,G 
\end{equation} 
with
\[ j_0 = \frac{3 L_0}{32\pi^2 \Gamma(1/3) m c^2} 
\left(\frac{ 2\pi m c}{\kappa q} \right)^{4/3}\]
\begin{equation} \label{GGG}
G = \gamma_1^{\alpha-1/3}\,Q^{5/3}(1)\,
Q^{-2/3}(2)\,U
\end{equation}
\[ U =
\int_{\gamma_1}^{\gamma_2} {\rm d}\gamma\;
\gamma^{-\alpha-2/3}\exp[-\nu/\nu_0/\gamma^2] 
\]
If $\nu \ll \nu_c$ we find $G \approx Q^{5/3}(1)\,
Q^{-2/3}(2)\,Q^{-1}(1/3)$.
To see the effect in the frequency range $\nu \approx \left<\gamma\,\right>^2
\nu_0$ we must evaluate the correct integral $U$.We introduce a new variable 
$s = \nu \nu_0^{-1}\gamma^{-2}$ an find :
\[ U =  
\left(\frac{\nu_0}{\nu}\right)^{\frac{\alpha}{2} - \frac{1}{6}}
\int_{s_2}^{s_1} {\rm d}s\; s^{\alpha/2 - 7/6}\;
\exp[-s] \]
\[ \nu_0 = \frac{3 B q}{4 \pi mc} \]
The integral $U$ can be expressed as the difference of two incomplete
Gamma-functions $\gamma(a,x)$ 
\[ U = \left(\frac{\nu_0}{\nu}\right)^{\frac{\alpha}{2} - \frac{1}{6}} 
\left(\gamma\left(\frac{\alpha}{2} - \frac{1}{6}, 
\frac{\nu}{\nu_0 \gamma_1^2}\right) 
-\gamma\left(\frac{\alpha}{2} - \frac{1}{6}, 
\frac{\nu}{\nu_0 \gamma_2^2}\right) \right) \]
or as the difference between two confluent hypergeometric
functions $M(a,b,z)$ (Kummer functions) resulting in a compact form for $G$
\[ 
G  =  2e^{-\frac{\nu}{\nu_c}}\,Q^{5/3}(1)\,Q^{-2/3}(2)\,
\left(\frac{\chi^{1/3-\alpha} \hat{Q}(\gamma_2) 
- \hat{Q}(\gamma_1)}{\frac{1}{3}-\alpha} \right)
\]
\[ \hat{Q}(\gamma) = \exp\left[-\nu/\nu_0\left(\gamma^{-2} - \left<\gamma\right>^{-2}\right)\right]
M\left(1,\frac{\alpha}{2}+\frac{5}{6}, \frac{\nu}
{\nu_0 \gamma^2}\right) \]
   \begin{figure}[htbp]
      \resizebox{\hsize}{!}{\includegraphics{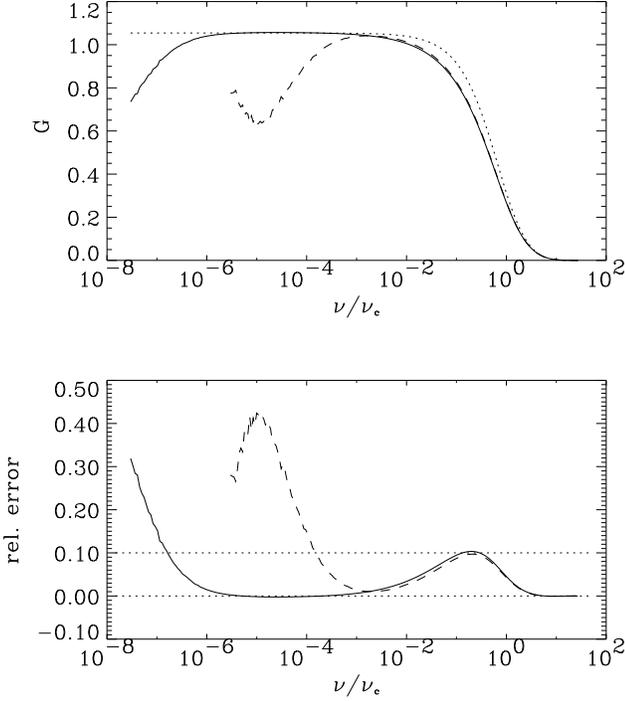}}
      \caption{Comparision of the approximation Eq. (\protect\ref{GGG})
        to the numerical evaluation of Eq. (\protect\ref{pf})
        for a truncated power-law of width $\chi = 3, \alpha = 2$
        averaged over all observers. The approximation is shown as a dotted 
        line and the numerical equivalents are plotted for $B = 10$ G and
        $\gamma = 1000$ (solid) and $\gamma = 100$ (dashed). The lower panel
        presents the relative error of the approximation. The dotted lines
        mark the 0\% and 10\% level
        .}
         \label{diff2}
   \end{figure}
   \begin{figure}[htbp]
      \resizebox{\hsize}{!}{\includegraphics{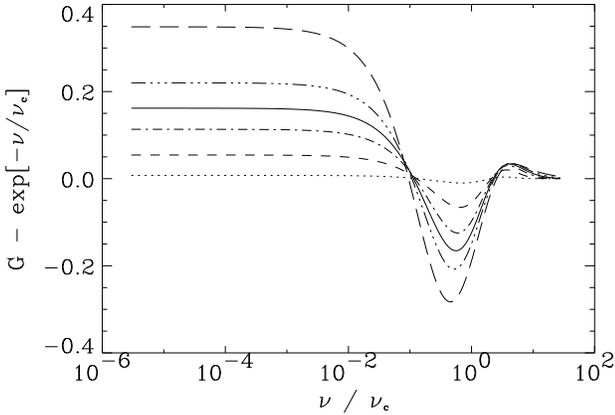}}
      \caption{Deviations of the synchrotron spectra of quasi-monoenergetic
          electrons from the strictly monoenergetic case. We took power-laws
          with $\alpha=2$ and plot 
          $G- \exp[-\nu/\nu_c]$ as defined in Eq. (\protect\ref{GGG}). 
          The deviations grow with $\chi = 1.5,3,5,7,10,20$.}
         \label{diff1}
   \end{figure}
The changes in the synchrotron spectra become significant when the
width of the distribution grows beyond a limit of say $\chi = 5$. In that case
the effect of the broader distribution exceeds the error introduced in the 
approximation Eq. (\ref{smpl}). The effect of the width on the approximative
spectra are shown in Fig. \ref{diff1} were the solid line correspond to 
$\chi = 7$.

For a homogeneous, spherical, optically thin source of radius $R$ 
we find for the received flux at a distance $d$
\begin{eqnarray}
S_\nu  & = & 2.01 \cdot 10^{-20}\left[\frac{R}{1{\rm AU}}\right]^3 
\left[\frac{d}{1{\rm kpc}}\right]^{-2} 
\left[\frac{B}{1{\rm G}}\right]^\frac{8}{3}
\left[\frac{\nu}{1{\rm GHz}}\right]^\frac{1}{3} 
\nonumber \\ & \times &
\left<\gamma\right>^{-\frac{5}{3}}  G 
\;{\rm erg}\;{\rm s}^{-1}
{\rm Hz}^{-1}\; {\rm cm}^{-2} 
\end{eqnarray}
with the coefficient $\kappa$ taken to be $1.2$ .
For completness we present 
the absorption coefficient for synchrotron self-ab\-sorp\-tion in the same
approximation as above for the emissivity. Inserting the mean emissivity from
Eq. (\ref{mem}) in Eq. (\ref{cross}) we get
\[ \alpha_\nu = \alpha_0 B^\frac{8}{3} \nu^{-\frac{5}{3}} 
\left<\gamma\right>^{-\frac{8}{3}}\,Q^{8/3}(1)\,Q^{-5/3}(2)\,J \]
\[ 
\alpha_0 = \frac{c^2 j_0}{2} = \frac{3 L_0}
{(8\pi)^2 \Gamma\left(\frac{1}{3}\right) m}
\left(\frac{4\pi m c}{3 q}\right)^{4/3}  \]
\[ J = (2 +\alpha)\gamma_1^{\alpha+2/3} \tilde{J} \]
\[ \tilde{J} =
\int_{\gamma_1}^{\gamma_2} {\rm d}\gamma\;
\gamma^{-\alpha-5/3}\exp\left[-\frac{\nu}{\nu_0\gamma^2}\right] 
\]
We will not discuss synchrotron self-absorption further in this approximation. 
%
\section{Aspects of radiative transfer in a homogeneous and spherical source
\label{flu}}
We assume that the strengh of the magnetic field $B$, the density of 
relativistic electrons $n(r)$ and the energy distribution of the 
electrons are constant throughout the source. As a consequence, the emissivity 
$\epsilon_\nu$ and the absorption coefficient $\alpha_\nu$ do not depend 
on the location inside the source. This simplifies the radiative transfer
so that we can get analytic expressions for the synchrotron flux at the 
surface. At first we find for the intensity 
\[ I_\nu = \frac{j_0}{\alpha_0} \nu^2 \frac{Q(2)\,\left<\gamma\right>}
{(2+\alpha)\,Q(1)\,\gamma_1} \frac{U}{\tilde{J}}
\left(1 - \exp\left[-\alpha_\nu s\right]\right) \]
if $s$ is the length of the path through the blob. Here we have again
assumed energy equipartition as in Sec.\ \ref{quas}. If we integrate
over the surface in the sky we find a total flux
\begin{eqnarray} 
  F_\nu & = & 2\pi\frac{\nu^2 }{c^2} \frac{Q(2)\,\left<\gamma\right>}
{(2+\alpha)\,Q(1)\,\gamma_1} \frac{U}{\tilde{J}} \times
  \nonumber \\ & & \label{fluss}
  \left(R^2 + \frac{R}{\alpha_\nu} e^{-2\alpha_\nu R}
  - \frac{e^{-\alpha_\nu R}}{\alpha_\nu^2}{\rm sinh}(\alpha_\nu R) \right) 
  \qquad .
\end{eqnarray} 
In the optically thick regime the flux is proportional to the
area $2\pi R^2$ and in the optically thin case we perform a Taylor expansion
and recover the simple form
\[ F_\nu \approx \frac{4\pi}{3}R^3 j_\nu \left(1 - \frac{3}{4}R \alpha_\nu
 \right) \]
to first order in in the optical depth $R\alpha_\nu$.
\section{Application to the galactic center source Sgr A$^*$ \label{SGR}}
The idea of monoenergetic relativistic electrons producing optically thin
synchrotron radiation has been applied to Sgr A$^*$ by Duschl \& Lesch (1994).
Later on, Narayan, Yi \& Mahadevan (1995) showed that thermal electrons
in an advection dominated accretion flow can lead to quite similiar
spectra, but an isothermal sphere seems to fit the data better.
In  B96
we discussed power-laws for isotropic electron distributions of limited width
with infinitely steep cut-offs and emphasized the importance of the absorption
mechanism for determining the source parameters. 
Here, we extend the distributions from the strict monoenergetic case discussed
above
to include more realistic power-laws with different cut-offs and
relativistic thermal distributions.
The spectra in  B96 were derived for a position angle of 
$\theta = 60\degr$ and in Sect.\ \ref{mol} we showed that this
is a good approximation to the average of all possible position angles.
Here we always take the average of all directions of magentic field lines. 
For a complete description of the source, we assume energy equipartition
in order to relate the magnetic field strength with the number density of electrons.
We calculate the spectral flux density according to Eq.\ (\ref{fluss}) and adopt
a distance of $8.5$ kpc to Sgr A$^*$.
%
   \begin{table}
      \caption{Physical parameters for the synchrotron source Sgr A$^*$}
         \label{smy}
         \begin{tabular}{rcccc}
            Model   &  $\left< E [{\rm MeV} ] \right>$ & $B$ [G] 
            & $R$ [$10^{13}$ cm] & $ n $  [$10^3 $cm$^{-3}$ ] \\
            \hline
            I &  140 & 10 & 1.25 &  18.0   \\
           II & 217 & 2.7 & 4.92 & 0.84 \\
          III & 222 & 3.0 & 4.86 & 1.01\\
           IV & 210 &  4.0 & 3.75 & 1.90 \\
            V & 202 & 4.0 & 3.50 & 1.97 \\
            \hline
         \end{tabular}
   \end{table}
The model {\sc v} in Tab.\ \ref{smy} is the best fit with a truncated power-law
distribution as discussed in Sec.\ \ref{quas} and B96.
\subsection{Power-laws with exponential cut-offs \label{expo}}
In B96 we introduced infinitely steep cut-offs for power-law distributions to 
achieve an easier understanding of the 
characteristics of synchrotron emission from quasi-mo\-no\-en\-er\-ge\-tic
electron distributions. They are not expected from acceleration mechanisms for 
electrons in a relativistic plasma. However, distribution functions with an 
exponential
cut-off at high energies and a rising power-law at low energies are expected for
equilibrium states (e.g. Schlickeiser 1984). If we think of an injection
mechanism at high energies like magnetic reconnection, we assume exponential 
decays at both ends of the distribution and a power-law in between. 
The distribution, normalized for a single electron, becomes
  \begin{equation}
  f(p) = K^{-1} p^\sigma \exp\left[ -\frac{p}{p_2} - \frac{p_1}{p}\right] 
  \end{equation}
  \[
  K = 2(p_1p_2)^{\frac{1+\sigma}{2}}
  {\rm K}_{1+\sigma}\left(2\sqrt{\frac{p_1}{p_2}}\right)  \quad .
  \]
The normalization is expressed in terms of  modified
Bessel functions ${\rm K}_z(x)$ of order $z$.
In contrast with power-laws covering a wide range of electron energies,
the spectral power $\sigma$ does not determine the emission spectrum and
is assumed to be $\sigma = -1$ in our case. The width $\chi = p_2/p_1$ is
defined in the same manner as in Sec.\ \ref{quas}.
The best possible model obtained with a single homogeneous blob of
such electrons is shown in Fig.\ \ref{therm}. The physical parameters are
given as model {\sc ii} in Tab.\ \ref{smy}.  In contrast to the spectra with
infinitely steep cut-offs, this model cannot account for the 
measured flux densities in the range $200 -1000 $GHz.
The power $\sigma$ and the width (here $\chi =3$) are unimportant for this part
of the spectrum. It is dominated by an exponential tail beyond $p_2$.
\subsection{Thermal distributions}
   \begin{figure}[htbp]
      \resizebox{\hsize}{!}{\includegraphics{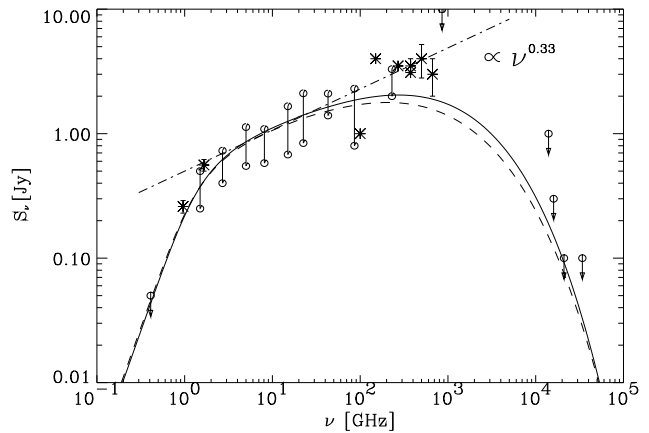}}
      \caption{Synchrotron spectrum of electron distributions with 
         exponential high energy tails. Both thermal electrons of 
         $T = 6\,10^{11} $K (solid line ; model {\sc iii} in 
         Tab.\protect\ref{smy}) and power-laws with exponential 
         cutoffs (dashed line; model {\sc ii})
         at high and low energies can hardly be distinguished. }
         \label{therm}
   \end{figure}
Another closely related spectrum is produced by relativistic thermal
electrons described by a distribution function
\[ f(p) = K p^2\exp\left[-\frac{\gamma}{\theta_e}\right] \qquad ,\]
as seen in Mahadevan et al. (1996) with the corresponding
normalisation $K$ to the number of electrons. The synchrotron spectrum beyond
the maxima at $1.2\gamma^3\nu_s$ 
is dominated by the exponential decay in the high energy band of the 
electron spectrum. This is also the case for the distributions
with exponential cut-offs discussed above. Obviously, both spatially 
homogeneous models suffer from
the same disease, since they can not explain the sub-mm data seen 
in Fig.\ref{therm}. 
The parameters for the thermal distribution are listed as model {\sc iii} 
in  Tab. \ref{smy}. 
\subsection{A Gauss-Profile for the electron momenta}
In the previous example we saw that an exponential high energy tail
of the distribution is in contradiction to the assumption of one single
homogeneous blob of optically thin synchrotron emission. 
A steeper tail is present in a Gaussian profile
   \begin{equation}
    f(p) = K^{-1}p^2\exp\left[-\frac{(p-p_1)^2}{p_2^2}\right]
   \end{equation}
with
   \[ K = \frac{p_2}{2}\sqrt{\pi}\left(2p_1^2 + p_2^2\right) \]
for the norm of the distribution.
   \begin{figure}[htbp]
      \resizebox{\hsize}{!}{\includegraphics{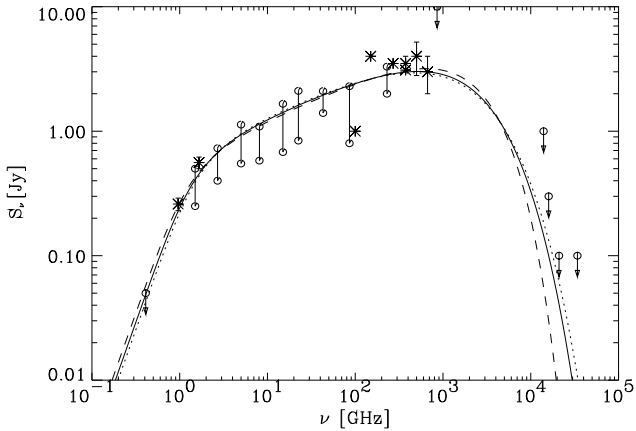}}
      \caption{Synchrotron flux from electrons with a gaussian profil 
               in momenta according to model {\sc iv} in Tab. 
               \protect\ref{smy}. The width is varied in the range
               $\chi = 0.2$ (dashed), $1.0$ (solid), $1.92$ (dotted) 
                and produce spectra with
                increasing flux densities at $2\,10^4$ GHz.
              }
         \label{gauss}
   \end{figure}  
We have included a phase space factor of $p^2$ which does not 
affect the resulting spectra very much. Again, the high energy tail is
the most important part necessary to explain the sub-mm measurements and yet
stay below the upper limits in the IR. 
We define the width of the Gaussian distribution $\chi$ as the FWHM $\Delta p$ 
of the pure gauss-profile relative to the mean momentum $p_1$ so that 
$\chi = \Delta p/p_1$ . The momentum $p_2$ is now
determined by the width $p_2 = 0.6\,\chi\,p_1$. Values greater than 2 are not 
reasonable for this definition of the width.
The obtained fit corresponds to model {\sc iv}
of Tab.\ref{smy}. We see in Fig. \ref{gauss} that this distribution
is sufficient to explain the radio spectrum of Sgr A$^*$ for a wide range of
widths up to $\chi = 1.92$.  
\subsection{A core-shell model for Sgr A$^*$}
Since distributions with an exponential high energy tail fail to explain the
full spectrum, using the assumption of isotropy and homogeneity of the source,
we can build up a two component source. There is good observational 
evidence for the existence of a central black hole of mass 
$M \approx 2.5\,10^6 M_\odot$ in the galactic center 
(Eckart \& Genzel 1996) which powers Sgr A$^*$. 
Accretion into this object implies a special radial structure suggested by
Narayan, Yi \& Mahadevan (1995). Our first step towards a radial
structure is a core-shell model. 

While the observational situation is far from clear yet, almost simultaneous 
observations of the spectrum of Sgr A* from cm to submm wavelengths (Falcke 
et al.\ 1997) seem to indicate the possibility of some execess submm flux in 
comparison to the flux predicted by the homogeneous model.

Our core-shell model consists of an optically thin 
extended source and a highly self-absorbed compact component. This is not 
the result of a hydrodynamical calculation, but the best description to explain
the observations. 
The extended part, component {\sc i},
dominates the spectrum up to $100$GHz and has its maximum at lower
frequencies compared with the one-component models. The compact component
{\sc ii} provides the sub-mm fluxes.
This picture is well distinguished from the radial non-uniform self-absorbed
source discussed by de Bruyn (1976) which is optically thick at any
wavelength, while our model is optically thin between $4 - 100$ GHz. 
   \begin{table}
      \caption{Two component model for the Synchrotron source Sgr A$^*$}
         \label{com}
         \begin{tabular}{rcccc}
            Comp.   &  $\left< E [{\rm MeV} ] \right>$ & $B$ [G] 
            & $R$ [$10^{13}$ cm] & $ n $  [$10^4 $cm$^{-3}$ ] \\
            \hline
            I &  155 & 2.0 & 5.5 & 0.064   \\
           II & 41.4  & 70  & 0.13  & 300 \\
            \hline
         \end{tabular}
   \end{table}
The drop of the electron energy inside
the compact core is an essential property of the model and is accompanied
by a significantly increased magnetic field in the core. 
The physical parameters are given in Tab. \ref{com}.
\section{Inverse-Compton Spectrum \label{ssc}}
The first consequence of
the interpretation of the radiospectrum as optically thin synchrotron radiation from
relativistic electrons is the synchrotron self-comptonization (SSC) process. 
In this process, only synchrotron photons are considered as  
the low energy radiation
scattered by relativistic electrons. The electron energy distribution 
$n_e(\gamma)$ is also obtained from the radiospectrum.    
From arguments on total cross sections for inverse Compton 
scattering (Blumenthal \& Gould 1970), we can estimate the emitted 
power due to comptonized synchrotron photons to be
 \[ P_{\rm Comp} = \frac{U_{\rm Ph}}{U_B} P_{\rm Sync} \qquad , \]
since synchrotron radiation can be considered as Compton scattering of the
virtual photons of the static magnetic field. The emitted power is scaled
by the relative energy densities of the photon field and the static
magnetic field. Taking an intrinsic synchrotron luminosity of $ 465
$L$_\odot$ for a gaussian electron distribution
and a source radius of $3.75\,10^{13}$ cm, we obtain an energy density of
$ U_{\rm Ph} = 3.4\,10^{-3}$ erg cm$^{-3}$, which falls short of the 
energy density of the magnetic field by more than two orders of magnitude. 
Thus we get a Inverse-Compton luminosity of
  \[ P_{\rm Comp} = 4.1\,10^{33} \left[ \frac{4.0 {\rm G}}{B}
  \right]^2 \left[\frac{3.75\,10^{13} {\rm cm}}{R}\right]^{-2} 
  \mbox{erg sec}^{-1} \quad.\]
   \begin{figure}[htbp]
      \resizebox{\hsize}{!}{\includegraphics{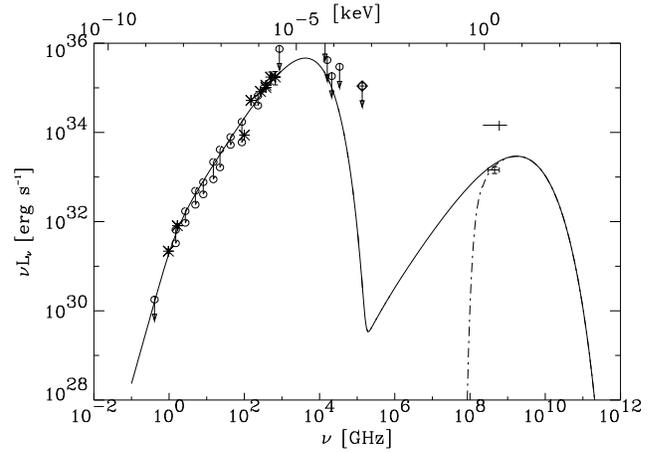}}
      \caption{The luminosity of Sgr A$^*$ for assumed distance of
               8.5 kpc from the radio to the soft X-ray regime. The model
               spectrum accounts for synchrontron and single 
               inverse-compton emission. The ROSAT and {\it Einstein}
               flux is included. The dashed line is the spectrum including
               absorption (see text).
              }
         \label{comp}
   \end{figure}
The spectrum of the scattered photons can be calculated using the
Thomson cross section $\sigma_T$, since the synchrotron photons are soft in 
the electron
rest-frame $¦\gamma h \nu \ll m_ec^2 $. For an isotropic electron and 
synchrotron-photon distribution (at least on average), 
the inverse-Compton flux is 
  \begin{equation} \label{compt}
   F_\nu = \frac{3}{16 \pi}\sigma_T c h \nu \int {\rm d}\gamma\;n_e(\gamma)
   \int {\rm d}\nu_s n_s(\nu_s) f(x) 
  \end{equation}
where we have used the distribution $f(x)$ (e.g Blumenthal \& Gould 1970) 
for single scattering events, defined as 
\[ f(x) = 2x\ln(x) +x +1 -2x^2 \qquad x = \frac{\nu}{4\nu_s \gamma^2} \]
and the index $s$ refers to the synchrotron photon distribution and frequency. 
The mean density of incident photons in a homogeneous spherical source of 
radius $R$ with
$\tau = \tau(R) = \alpha_\nu R$  and the source function $S_{\nu_s}$ is given by
\[ n_s(\nu_s) = \frac{4\pi}{c h \nu_s} S_{\nu_s}\left( 1 - \frac{1}{\tau}
+ \frac{1 - e^{-2\tau}}{2\tau^2}\right) \qquad . \]
For most distributions, the integrals in  Eq. (\ref{compt}) can not be evaluated
analytically. The calculated spectrum shown in Fig.\ \ref{comp} is
produced by a gaussian electron distribution according to model {\sc IV}
of Tab.\ref{smy} .
The measured flux density in the $0.8-2.5$ keV range reported by
Predehl \& Tr\"umper (1994) for ROSAT-PSPC without correction for absorption
is included in Fig. \ref{comp}. We take the upper limit for the ionized hydrogen
column density towards Sgr A$^*$ (B96)
$N_{\rm H} = 2.2\ 10^{21} $ cm$^{-2}$ and obtain a cut-off due to
photoelectric absorption (Morrison \& McCammon 1983) at approximatly
1 keV. SSC and external absorption with the
predicted column density from the turn-over at 1 GHz 
can account for the ROSAT measurement. Other measurments with {\it Einstein}
(Watson et al. 1981) and ART-P (Pavlinsky et al. 1994) are not consistent
with this interpretation.  The assumed $N_{\rm H}$ column density 
is much too low to account for the visual extinction of approximately
$30^{\rm mag}$ corresponing to a canonical hydrogen column density of
$N_{\rm H} = 6\ 10^{21}$ cm$^{-2}$ towards the galatic center.
%
\section{Summary \label{conc}}
Synchrotron radiation by quasi-monoenergetic electrons is
considered as an explanation of the radio spectrum of the 
galactic center source Sgr A$^*$.
Therefore, we developed a detailed treatment of the emission process
of single electrons in a homogeneous magnetic field to obtain a correct
synchrotron spectrum even for mildly relativistic electron energies.
We showed that the emissivity of a single electron is modified due to the
relativistic motion of the electrons with respect to the observer
compared with the standard theory. This effect is removed when a
stationary and isotropic distribution is considered as shown in   
Appendix B.   
The spectra of monoenergetic electrons are integrated for various electron
distributions covering a thin shell in momentum space. These 
quasi-monoenergetic electrons in a homogeneous and isotropic source
provide a fairly good fit to the time averaged radiospectrum of Sgr A$^*$.
The resulting self-compotonized X-rays are sufficient to explain the
ROSAT-observations with a very low hydrogen column density. This hydrogen 
column density is consistent with the upper limit of ionized hydrogen in the 
vicinity of Sgr A$^*$ (B96) infered from free-free absorption. Additional 
material is requiered to account for the visual extinction. This may also
lead to enhanced absorption in the soft X-ray range and the synchrotron 
self-comptonization would not be sufficient to explain the flux measured by
ROSAT.  
 
Inverted radiospectra showing a synchrotron flux proportional to
$\nu^{0.3}$  are also reported for the centers of M81 (Reuter\- \&\- 
Lesch 1996),
NGC 1068 (Wittkowski et al. 1997) the archetypal Seyfert 2 galaxy and 
M 104 (Jauch \& Duschl, in preparation).
The spectra of all these sources can be interpreted as due to optically 
thin synchrotron radiation from quasi-monoenergetic electrons. 
This suggest a new 
common feature for the centers of normal and active galaxies. The differences
in the radio regime arise from the attainable electron energies and the
source size. 
\begin{acknowledgements}
We thank the referee, R.\ Narayan, and R.\ Mahadevan for helpful remarks
which improved the paper, P.G.\ Mezger for interesting discussions of the
subject, and R.\ Auer for carefully reading the maunscript. This work was 
supported by the {\it Deutsche Forschungsgemeinschaft\/} through SFB 328.
\end{acknowledgements}
%
\appendix 
\section{The Radiation Field at the Observer \label{dett}}
The electric field at the position of the observer can be expressed in terms of
source properties at the retarded time (Jackson 1962)
   \begin{eqnarray} \vec{E} & = & \frac{q}{cR}\vec{g} \\
   \vec{g}(t) & = & \left.\frac{\vec{n}\times(\vec{n}-\vec{\beta})\times
   \dot{\vec{\beta}}}{(1-\vec{n}\cdot\vec{\beta})^3}\right|_{\rm ret.} 
   \qquad . \label{retg}
   \end{eqnarray}
We take the viewpoint of an observer at an angle $\theta$ between the 
line of sight and the magnetic field lines.
Assuming a large distance between the source and the observer 
we approximate the distance at the retarded time  $t = \tau + R(\tau)/c$ as
   \begin{equation} \label{abst}
   R(\tau)  \approx R - \vec{n}\vec{r}(\tau) \qquad .
   \end{equation}
The vector 
\begin{equation}
   \vec{r}(\tau) = \frac{\beta c}{\omega_s} 
   \left(\begin{array}{c}  \sin\alpha\cos \phi \\
   \sin\alpha\,\cos \theta\,\sin\phi - \tau\omega_s\sin\theta\,\cos\alpha \\
   \sin\alpha\,\sin \theta\,\sin\phi + \tau\omega_s\cos\theta\,\cos\alpha \end{array}\right) 
\end{equation} 
describes the path of the spiraling electron 
with the azimuth angle $\phi = \omega_s \tau$ 
inside the source and $\vec{n} = (0,0,1)$ the line of sight in a coordinate
system with the observer in the $z$-direction.
If we take $\theta$ to be a constant for sufficiently long times, it follows
that $\vec{n}\vec{r}(\tau)$ becomes a periodic function in $\tau$.
This forces us to expand the electric field  as was 
previously stated by Shu (1991). Thus the
spectrum of a single electron decomposes into discrete emission lines with
Fourier coefficients given by
   \begin{equation}
   \vec{g}_m = \frac{\omega}{2\pi}
   \int_{-\frac{\pi}{\omega}}^{\frac{\pi}{\omega}}
   {\rm d}t\; \vec{g}(t)\exp[-im\omega t] 
   \label{gem}
   \end{equation}
and the time average of the observed power is provided by Parseval's theorem 
   \begin{equation}\label{pow}
   P(\Omega) = \nu\int_0^{\frac{1}{\nu}}{\rm d} t\; P(\Omega,t)
    = \frac{q^2}{2\pi c} 
   \sum_{m=1}^\infty \left|\vec{g}_m\right| 
   \end{equation}
and the symmetry of $\vec{g}_m$ with respect to $m \rightarrow -m$.
We trace back the integral to the retarded time and, following the derivation
given by Bekefi (1966), we obtain
   \begin{equation}
   -im\omega t \approx -im(\omega_s \tau - \psi \sin(\omega_s\tau))
   \end{equation}
 for the argument of the exponential function.
This provides us with a parameter reflecting the geometry, $\psi$, and connects
the frequency of the spiral with the frequency measured by the 
observer $\omega$. 
The derivative of the light cone conditon 
\begin{equation}
{\rm d}t = {\rm d}\tau \left( 1 +\frac{{\rm d}R(\tau)}{c\, {\rm d} \tau}\right) 
\end{equation}
allows a substitution of the integration variable $t$ by $\tau$ in the same way as 
Jackson (1962 Cap. 14.5).
Using approximation (\ref{abst}) and the identity
\begin{equation} 
  \frac{\rm d}{{\rm d}\tau} \frac{\vec{n}\times \left[ \vec{n}\times\vec{\beta}
  \right]}{1-\vec{n}\vec{\beta}} =  (1-\vec{n}\vec{\beta})\, \vec{g}(\tau) 
\end{equation}
we perform a partial integration of (\ref{gem}) and get 
\begin{equation}
  \vec{g}_m = -\frac{im\omega^2}{2\pi}
 \int_{-\tau_1}^{\tau_1} {\rm d}\tau\, \vec{g}_1
  \exp[-im\omega t] 
\end{equation}
with
\begin{eqnarray} 
  \vec{g}_1 & = &   \left(\begin{array}{c}   \sin \alpha \sin \phi \\
   -\sin\alpha\, \cos \theta\,\cos\phi + \sin\theta\,\cos\alpha \\
   0  \end{array}\right) \\
  t & = & ( \tau(1-\beta\cos\alpha\,\cos\theta) - \beta \omega_s^{-1}
  \sin\alpha\,\sin\theta,\sin(\omega_s\tau)) \nonumber
\end{eqnarray}  
The contribution from the boundary in the partial integration vanishes if the
argument is periodic  with period $2\pi\,\omega^{-1}$ which determines the
frequency $\omega = (1-\beta\cos\theta\,\cos\alpha)^{-1}\omega_s$ seen by the
observer. Changing the variable of integration from $\tau$ to $\phi$ we get
\begin{equation} \label{apg}
  \vec{g}_m = \frac{im\omega_s\beta}{2\pi \xi^2}\int_{-\pi}^{\pi} 
  {\rm d}\phi\, \vec{g}_1 
  \exp[-im(\phi - \psi\sin\phi)]
\end{equation}
with $\psi$ and $\xi$ defined in (\ref{freq}). The argument of the exp-function
in (\ref{apg}) is antisymmetric in $\phi \rightarrow - \phi$ and the 
$x$-component of $\vec{g}_1$ is also antisymmetric. The $x$-component of
$\vec{g}_m$ is given by the symmetric part of the integrand
as 
\begin{eqnarray}
& & \beta \frac{m\omega_s}{2 \pi \xi^2}\,\sin\alpha \; \int_0^\pi {\rm d}\phi\,
\left[\cos((m-1)\phi - m\psi\sin\phi) \right.\nonumber \\
& - & \left.\cos((m+1)\phi - m\psi\sin\phi)\right] 
\end{eqnarray}
Taking the integral representation for Bessel functions of integer order
(e.g. Abramowitz \& Stegun 1972 [9.1.21])
\begin{equation}
 \pi J_m(z)  =  \int_0^\pi {\rm d}\phi \cos(z\sin\phi - m \phi)
\end{equation}
and the recurrence relation
$J_{m-1}(z) - J_{m+1}(z) = 2 J'_m(z)$ the $x$-component of $\vec{g}_m$ becomes
\begin{equation} \label{gx}
\frac{m\omega_s}{\xi^2}\beta\sin\alpha  J'_m(m\psi)
\end{equation}
The $y$-component of $\vec{g}_1$ is symmetric and the part contributing to the
$y$-component of $\vec{g}_m$ is
\begin{eqnarray}
& & i\frac{m\omega_s}{2\xi^2}\beta \left[2 \cos\alpha \sin\theta  J_m(m\psi)
 \right. \nonumber \\
& - &  \left. \sin\alpha \cos\theta (J_{m-1}(m\psi) + J_{m+1}(m\psi)) \right]
\end{eqnarray}
With an additional recurrence relation $J_{m-1}(z) + J_{m+1}(z) = 2m z^{-1} J_m(z)$ 
the $y$-component of $\vec{g}_m$ can be rewritten as
\begin{equation} \label{gy}
 i\frac{m\omega_s}{\xi^2} \frac{\beta\cos\alpha - \cos\theta}
 {\sin\theta}  J_m(m\psi)
\end{equation}
The two representations (\ref{gx}) and (\ref{gy}) for the two components of the 
electric field are presented in Eq. (\ref{sqr}).
\section{The Volume-Integral of Recieved Power \label{phas}}
   \begin{figure}[htbp]
      \resizebox{\hsize}{!}{\includegraphics{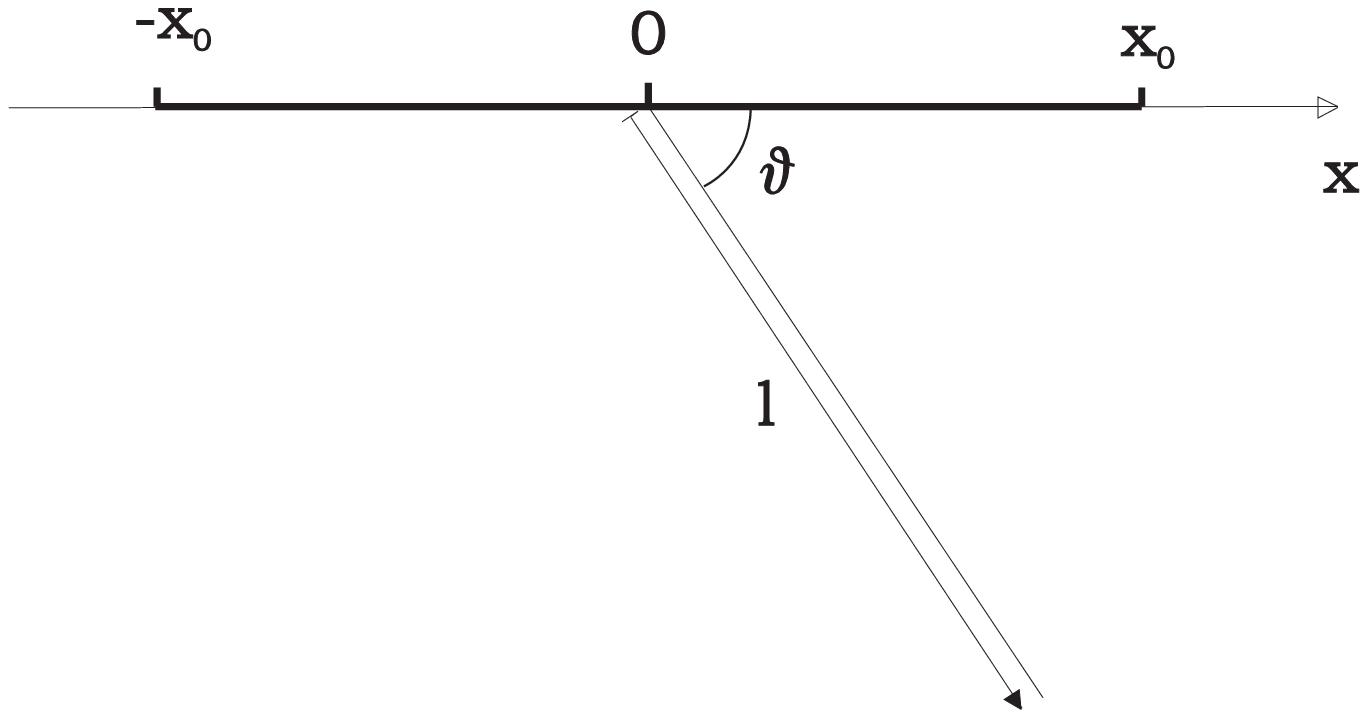}}
   \caption{Geometry of the the source region and choice of coordinates.} 
   \end{figure}
%
\subsection{Geometry and Coordinates}
We consider the emission of synchrotron radiation by relativistically
moving electrons in a homogeneous magnetic field which is inclined to the
observer by an angle $\vartheta$. The source is located far from the 
observer and is unresolved, so that the recieved power does not depend
on the position of the emitting electron inside the source.
Due to the motion of an individual electron along the magnetic field 
with velocity $v_\|=c \beta_\|$, 
this electron is seen by the observer for a time
$\Delta t = v_\| x^{-1} (1\pm|\beta_\||\cos \vartheta)$ if $x$ is the
length of the electron's path along the magnetic field during the
observation. The time interval is shortened for electrons approaching
the observer and prolonged for those which are receding. 
For calculating the observed power,
Scheuer (1968) showed that the received power of an individual electron 
is reduced by a factor $(1\pm\beta_\|\cos \vartheta)$ since it contributes
to the total power only for a velocity-dependent time interval.
When dealing with a distribution function in phase space discribing the
electrons in the source we show here that this argument must be based on the 
invariance of the distribution function to get the same result.
In order to simplify the discussion we consider the source to be a box of length
$2x_0$ with the magnetic field directed along the $x-$axis and the
centre of the coordinate system centered in the box. Suppose we have two
fluxes of electrons all of the same energy. 
One moving in the negative $x-$direction with velocity $-v_\|$ and the
other in the positive direction with $v_\|$. Synchrotron radiation should
only be emitted while the electrons are inside the box and the distance between
two neighbouring electrons moving in the same direction is $x_0/N$.
The distribution of electrons can be written as
  \begin{eqnarray}
  f(x) & = & \Theta (x_0 - |x|)\left[\sum_n \delta\left(x-n\frac{x_0}{N} + 
  c\beta_\|t \right) + \right.\nonumber \\ & + &
  \left. \sum_m\delta\left(x+m\frac{x_0}{N} - c\beta_\|t
  \right)\right]
  \end{eqnarray}
For each emitting electron it takes a time $\Delta t_e$ for the radiation
to arrive at the position of the observer, given by
  \begin{equation}
  \Delta t_e = c^{-1}\left( l-x\cos\vartheta \right) \quad ,
  \end{equation} 
which depends on the distance between source and observer $l$ which is omitted
in the following analysis. 
\subsection{Normalizing the Distribution}
In general the distribution function and its integral in phase space is
an invariant due to changes in the variables. This can be expressed as
\[ \int_{\Delta t}{\rm d}t\; \int {\rm d}^3 r\; {\rm d}^3 p\; 
\hat{f}(\vec{r},\vec{p},t) = 4N \Delta t \]
with the knowledge of the energy equation $E^2 = p^2c^2 + m_e^2 c^4$. We have 
$4N$ electrons on average in the volume of interest and the complete 
distribution function of our problem reads
  \begin{eqnarray} 
  \hat{f} & = & \left[\delta^3(\vec{p}-\vec{p}_0)+\delta^3(\vec{p}+\vec{p}_0)\right]
  \Theta(x_0 -  |x|) \nonumber \\ & & 
  \delta(y-y_0)\delta(z-z_0)
  \sum_n\delta\left(x-n\frac{x_0}{N} -\frac{p_x}{m_e}t \right)
  \end{eqnarray}
where we used the momentum $\vec{p}_0 = (v_\|m_e , 0, 0)$. Performing the 
simple integrals gives us 
\[ 4N \Delta t = \int_{\Delta t}{\rm d}t\; \int{\rm d}x\; \int {\rm d}p_x\;
\tilde{f}(x,p_x,t) \qquad .\] 
If we want to express the integral in terms of the arrival time of signals at
the observer $\tilde{t} = t + \frac{x}{c}\cos\vartheta$, we have to change the
integration variables, resulting in a factor
\[ \frac{\partial(\tilde{x},\tilde{p}_x,\tilde{t})}{\partial(x,p_x,t)} =
1+\frac{p_x}{m_ec}\cos\vartheta \]
with $\tilde{x} = x$ and $\tilde{p}_x = p_x$. So, whenever we calculate 
mean values of observables $P(x,p_x,t)$ in terms of $x,\tilde{t}$, we have to
take
  \begin{eqnarray}  \label{PA} 
  & & P(\tilde{t})  =  \nonumber \\ & &
  \int_{-x_0}^{x_0}{\rm d} x\, 
  \left[P_+\sum_n\xi_+\,\delta\left(x-n\frac{x_0}{N} + 
  c\beta_\|\left(\tilde{t} + \frac{x}{c}\cos\vartheta\right) \right)\right. 
  \nonumber \\ & & +  \left.
  P_-\,\sum_m\xi_-\,\delta\left(x+m\frac{x_0}{N} - c\beta_\|
  \left(\tilde{t} + \frac{x}{c}\cos\vartheta\right)\right)\right] 
  \end{eqnarray}
\begin{eqnarray}
\xi_+ & = 1 + \beta_\| \cos \vartheta \qquad P_+ & = P(x,+p_0) \nonumber \\
\xi_- & = 1 - \beta_\| \cos \vartheta \qquad  P_- & = P(x,-p_0)
\end{eqnarray}
\subsection{Mean observed synchrotron power}
If we know the received power for the two different types of electrons to be
$P_+,P_-$ we can integrate to get the total received power. 
If we collect terms containing $x$ in the argument of the $\delta$-functions 
in (\ref{PA}), we find
  \begin{eqnarray}
  P(\tilde{t}) & = & \int_{-x_0}^{x_0} {\rm d} x\;
  \left[P_+\,\xi_+\;\sum_n \delta\left(\xi_+x-n\frac{x_0}{N} + 
  c\beta_\|\tilde{t} \right) + \right. \nonumber \\ & + & \label{Pt} \left.
  P_-\,\xi_-\;\sum_m\delta\left(\xi_-x+m\frac{x_0}{N} - c\beta_\|\tilde{t}\right)\right]
  \end{eqnarray}
for the total power recieved.
To evaluate the spatial integral we transform the argument of the
$\delta-$functions 
  \begin{eqnarray}
  P & = & P_+\int_{-x_0}^{x_0} {\rm d} x\;
  \sum_n \delta\left(x-n\frac{x_0}{\xi_+N} + 
  c\frac{\beta_\|\tilde{t}}{\xi_+} \right) + \nonumber \\ 
  & + & P_-\int _{-x_0}^{x_0}
  \sum_m\delta\left(x+m\frac{x_0}{\xi_-N} - 
  c\frac{\beta_\|\tilde{t}}{\xi_-}\right)
  \end{eqnarray}
according to $\int \delta(f(x))$d$x = \sum_n\delta(x-x_n) \| f'(x_n)\|^{-1}$.
The conditions under which the electrons, which are counted by $n$ and $m$,
contribute to the integrals is
\begin{eqnarray}
\left(\frac{c}{x_0}\beta_\| \tilde{t} - \xi_+\right)N & \le n < &
\left(\frac{c}{x_0}\beta_\| \tilde{t} + \xi_+\right)N \nonumber\\
\left(\frac{c}{x_0}\beta_\| \tilde{t} - \xi_-\right)N & \le m < &
\left(\frac{c}{x_0}\beta_\| \tilde{t} + \xi_-\right)N \nonumber \\
\label{con} \Delta n = 2 \xi_+ N & \& & \Delta m = 2 \xi_- N 
\end{eqnarray}
and we conclude that, for any instance of time,
the number of electrons contributing to the integrals is 
$2N\xi_+$ and $2N\xi_-$ respectively. In this way the argument given above
for the different time intervals is recovered. We see that the total
received power is changed by this counting argument only because the change
in the time as a variable $t \rightarrow \tilde{t}$ is compensated by the
invariance of the integral of the distribution function in $\Gamma$-space.
This results in
\begin{equation}\label{PP}
  P = 2N(1+\beta_\|\cos\vartheta) P_+ + 2N(1-\beta_\|\cos\vartheta) P_- 
\end{equation}
The results can alternatively be derived by multipling the number of observed
electrons of each kind given in (\ref{con}) by the power recieved from these
electrons.
In this way, the result based on counting individual electrons (Scheuer 1968, 
Rybicky \& Lightman 1979), can be recovered.
%

\end{document}